\documentclass{sigchi-ext}
\usepackage[T1]{fontenc}
\usepackage{textcomp}
\usepackage[scaled=.92]{helvet} % for proper fonts
\usepackage{graphicx} % for EPS use the graphics package instead
\usepackage{balance}  % for useful for balancing the last columns
\usepackage{booktabs} % for pretty table rules
\usepackage{ccicons}  % for Creative Commons citation icons
\usepackage{ragged2e} % for tighter hyphenation

\def\plaintitle{Some HCI Priorities for GDPR-Compliant Machine Learning} 
\def\emptyauthor{}
\def\plainkeywords{GDPR; machine learning; anti-discrimination; data protection; data protection by design; algorithmic accountability}

\numberofauthors{3}
\author{%
  \alignauthor{%
    \textbf{Michael Veale}\\
    \affaddr{University College London} \\
    \affaddr{London, United Kingdom} \\
    \email{m.veale@ucl.ac.uk} }\alignauthor{%
    \textbf{Reuben Binns}\\
    \textbf{Max Van Kleek}\\    
    \affaddr{University of Oxford}\\
    \affaddr{Oxford, United Kingdom}\\
    \email{reuben.binns@cs.ox.ac.uk\\ emax@cs.ox.ac.uk}
    }}
\copyrightinfo{ \emph{The General Data Protection Regulation: An Opportunity for the CHI Community? (CHI-GDPR 2018)}, Workshop at ACM CHI'18, 22 April 2018, Montr\'{e}al, Canada}
\usepackage{balance}    % useful for balancing the last columns
\definecolor{linkColor}{RGB}{6,125,233}

\hypersetup{%
  pdftitle={\plaintitle},
  pdfauthor={\emptyauthor},
  pdfkeywords={\plainkeywords},
  bookmarksnumbered,
  pdfstartview={FitH},
  colorlinks,
  citecolor=black,
  filecolor=black,
  linkcolor=black,
  urlcolor=linkColor,
  breaklinks=true,
}

\usepackage{microtype}
\title{Some HCI Priorities for GDPR-Compliant Machine Learning}

\begin{document}
% \maketitle
%\titlenote{Produces the permission block, and copyright information}

%\author{Michael Veale}
%\orcid{0000-0002-2342-8785}
%\affiliation{%
% \institution{University College London}
%% \streetaddress{36–38 Fitzroy Square}
% \city{London}
% \country{United Kingdom}
%}
%\email{m.veale@ucl.ac.uk}
%
%\author{Reuben Binns}
%\affiliation{%
% \institution{University of Oxford}
%% \streetaddress{36–38 Fitzroy Square}
% \city{Oxford}
% \country{United Kingdom}
%}
%\email{reuben.binns@cs.ox.ac.uk}
%
%\author{Max Van Kleek}
%\affiliation{%
% \institution{University of Oxford}
%% \streetaddress{36–38 Fitzroy Square}
% \city{Oxford}
% \country{United Kingdom}
%}
%\email{max.van.kleek@cs.ox.ac.uk}

%
% The code below should be generated by the tool at
% http://dl.acm.org/ccs.cfm
% Please copy and paste the code instead of the example below.
%

%\keywords{GDPR, machine learning, anti-discrimination, data protection, data protection by design, algorithmic accountability}

\maketitle

\begin{abstract}
In this short paper, we consider the roles of HCI in enabling the better governance of consequential machine learning systems using the rights and obligations laid out in the recent 2016 EU General Data Protection Regulation (GDPR)---a law which involves heavy interaction with people and systems. Focussing on those areas that relate to algorithmic systems in society, we propose roles for HCI in legal contexts in relation to fairness, bias and discrimination; data protection by design; data protection impact assessments; transparency and explanations; the mitigation and understanding of automation bias; and the communication of envisaged consequences of processing. 
\end{abstract}

\section{Introduction}
%\sloppy
The 2016 EU General Data Protection Regulation (GDPR) is making waves. With all personal data relating to EU residents or processed by EU companies within scope, it seeks to strengthen the rights of data subjects and the obligations of data controllers (see definitions in the box overleaf) in an increasingly data-laden society, newly underpinned with an overarching obligation of data controller accountability as well as hefty maximum fines. Its articles introduce new provisions and formalise existing rights clarified by the European Court of Justice (the Court), such as the ``right to be forgotten'', as well as strengthening those already present in the 1995 Data Protection Directive (DPD).

The GDPR has been turned to by scholars and activists as a tool for ``algorithmic accountability'' in a society where machine learning (ML) seems to be increasingly important. Machine learning models---statistical systems which use data to improve their performance on particular tasks---are the approach of choice to generate value from the `data exhaust' of digitised human activities.
Critics, however, have framed ML as powerful, opaque, and with potential to endanger privacy~\cite{BarocasNiss2014}, equality~\cite{Custers2012x} and autonomy~\cite{Hildebrandt2008}. While the GDPR is intended to govern personal data rather than ML, there are a range of included rights and obligations which might be useful to exert control over algorithmic systems~\cite{edwardsveale}.

\marginpar{%
  \vspace{-150pt} \fbox{%
    \begin{minipage}{0.925\marginparwidth}
      \textbf{Data Subjects \& Controllers} \\
 EU DP law applies whenever personal data is processed either in the Union, or outside the Union relating to an EU resident. Personal data is defined by how much it can render somebody identifiable---going beyond email, phone number, etc to include dynamic IP addresses, browser fingerprints or smart meter readings. The individual data relates to is called the \textit{data subject}. The organisation(s) who determine `the purposes and means of the processing of personal data' are \textit{data controllers}. Data subjects have rights over personal data, such as rights of access, erasure, objection to processing, and portability of data elsewhere. Data controllers are subject to a range of obligations, such as ensuring confidentiality, notifying if data is breached, and undertaking risk assessments. Additionally, they must only process data where they have a legal ground---such as consent---to do so, for a specified and limited purpose, and a limited period of storage.
    \end{minipage}}\label{sec:sidebar} }

Given that GDPR rights involve both individual data-subjects and data controllers (see sidebar) interfacing with computers in a wide variety of contexts, it strongly implicates another abbreviation readers will likely find familiar: Human--Computer Interaction (HCI). In this short paper, we outline, non-exhaustively of course, some of the crossovers between the GDPR provisions, HCI and ML that appear most salient and pressing given current legal, social, and technical debates. We group these in two broad categories: those which primarily concern the building and training of models \emph{before deployment}, and those which primarily concern the \emph{post-deployment application} of models to data subjects in particular situations.
    
\section{HCI, GDPR and Model Training}

An increasing proportion of collected personal data\footnote{Note that the GDPR defines personal data broadly---including things like dynamic IP addresses and home energy data---as opposed to the predominantly American notion of personally identifiable information (PII)~\cite{schwartz2011pii}.} is used to train machine learning systems, which are in turn used to make or support decisions in a variety of fields. As model training with personal data is considered data processing (assuming data is not solidly `anonymised'), the GDPR does govern it to a varying degree. In this section, we consider to what extent HCI might play a role in promoting the governance of model training under the GDPR.

\subsection{Fairness, discrimination and `special category' data}

Interest in unfair and/or illegal data-driven discrimination has concerned researchers, journalists, pundits and policy-makers~\cite{gandy2009,barocas2016big}, particularly as the ease of transforming seemingly non-sensitive data into potentially damaging, private insights has become clear~\cite{Calders1:2012}. Most focus on how to govern data (both in Europe and elsewhere broadly~\cite{greenleaf2017Europe}) has been centred on data protection, which is not an anti-discrimination law and does not feature anti-discrimination as a core concept. Yet the GDPR does contain provisions which concern particularly sensitive attributes of data.

Several ``special'' types of data are given higher protection in the GDPR. The 1995 Data Protection Directive (art 8) prohibits processing of data \emph{revealing} \textbf{racial or ethnic origin}, \textbf{political opinions}, \textbf{religious or philosophical beliefs} and \textbf{trade-union membership}, in addition to data concerning \textbf{health} or \textbf{sex life}. The GDPR (art 9(1)) adds \textbf{genetic} and \textbf{biometric} data (the latter for the purposes of identification), as well as clarifying sex life includes orientation, to create 8 `special categories' of data. This list is similar, but not identical, to the `protected characteristics' in many international anti-discrimination laws. Compared to the UK's Equality Act 2010, the GDPR omits age, sex and marital status but includes political opinions, trade union membership, and health data more broadly. 

The collection, inference and processing of special category data triggers both specific provisions (e.g. arts 9, 22) and specific responsibilities (e.g. Data Protection Impact Assessments, art 35 and below), as well as generally heightening the level of risk of processing and therefore the general responsibilities of a controller (art 24). Perhaps the most important difference is that data controllers cannot rely on their own \emph{legitimate interests} to justify the processing of special category data, which usually will mean they will have to seek explicit, specified consent for the type of processing they intend---which they may not have done for their original data, and may not be built into their legal data collection model.

Given that inferred special category data is also characterised as special category data~\cite{vealeedwardsa29}, there are important questions around how both controllers and regulators recognise that such inference is or might be happening. Naturally, if a data controller trains a supervised model for the purpose of inferring a special category of data, this is quite a simple task (as long as they are honest about it). Yet when they are using latent characteristics, such as through principal components analysis, or features that are embedded within a machine learning model, this becomes more challenging. In particular it has been repeatedly demonstrated that biases connected to special category data can appear in trained systems even where those special categories are not present in the datasets being used~\cite{Calders1:2012}.

The difficulty of this task is heightened by how the controller is unlikely to possess `ground truth' special category data in order to assess what it is they are picking up. HCI might play an important role here in establishing what has been described as `exploratory fairness analysis'~\cite{vealebinns2017}. The task is to understand potential patterns of discrimination, or to identify certain unexpected but sensitive clusters, with only partial additional information abut the participants. A similar proposal (and prototype of) a visual system, albeit one assuming full information, has been proposed by discrimination-aware data mining researchers concerned that the formal statistical criteria for non-discrimination established by researchers may not connect with ideas of fairness in practice~\cite{berendt2012exploring,berendt2014better}. If we do indeed also know unfairness when we see it, exploratory visual analysis may be a useful tool. A linked set of discussions have been occurring in the information visualisation community around desirable characteristics of feminist data visualisation, which connects feminist principles around marginalisation and dominance in the production of knowledge to information design~\cite{d2016feminist}. Finally, visual tools which help identify misleading patterns in data, such as instances of Simpson's paradox (e.g. \cite{matejka2017same}), may prove useful in confirming apparent disparities between groups. Building and testing interfaces which help identify sensitive potential correlations and ask critical questions around bias and discrimination in the data is an important prerequisite to rigorously meeting requirements in the GDPR.

% Special categories feature in a more minor way in other aspects of the GDPR. Firstly, there are slightly greater hurdles to overcoming the prohibition on solely automated, significant decisions if they are taken on the basis of special category data \cite{edwardsveale}. Secondly, there is a discussion of using fairness-aware data mining (albeit in a non-binding recital that precedes the main text of the GDPR,  where ``\emph{the controller should use appropriate mathematical or statistical procedures for the profiling, implement technical and organisational measures appropriate [...] that [prevent], inter alia, discriminatory effects on natural persons on the basis of [special categories of data], or that result in measures having such an effect.}'' However, like much of the widely misunderstood `right to an explanation' that many believe to exist in a strong form in the GDPR, this highly political text exists only in the long grass of the non-binding recitals that precede the main articles \cite{wachter2017right,edwardsveale}, and therefore it is unlikely to be considered a mandatory requirement by regulators or the courts, except perhaps in the most sensitive, high-stakes situations. Thirdly, there is a requirement to inform data subjects that special categories of data are being derived from the non-special category data that relate to them. Finally, processing special category data will make the need to carry out a data protection impact assessment (DPIA) before processing more likely to be required.

\subsection{Upstream provisions: Data Protection by Design (DPbD) and Data Protection Impact Assessments (DPIAs)}

The GDPR contains several provisions intended to move considerations of risks to data subjects' rights and freedoms upstream into the hands of designers. Data Protection by Design (DPbD), a close cousin of privacy by design, is a requirement under the GDPR and means that controllers should use organisational and technical measures to imbue their products and processes with data protection principles~\cite{bygrave2017data}. Data Protection Impact Assessments (DPIAs) have a similar motivation~\cite{binns2017data}. Whenever controllers have reason to believe that a processing activity brings high risks, they must undertake continuous, documented analysis of these, as well as any measures they are taking to mitigate.

The holistic nature of both DPbD and DPIAs is emphasised in both the legal text and recent guidance. These are creative processes mixing anticipation and foresight with best practice and documentation. Some HCI research has already addressed this in particular. Luger et al.~\cite{luger2015playing} use \textit{ideation cards} to engage designers with regulation and co-produce ``data protection heuristics''.\footnote{The cards are downloadable at \url{https://perma.cc/3VBQ-VVPQ}.} Whether DPIA aides can be built into existing systems and software in a user-centric way is an important area for future exploration.

Furthermore, many risks within data, such as bias, poor representation, or the picking up of private features, may be unknown to data controllers. Identifying these is the point of a DPIA, but subtle issues are unlikely to leap out of the page. In times like this, it has been suggested that a shared knowledgebase~\cite{vealebinns2017} could be a useful resource, where researchers and data controllers (or their modelling staff) could log risks and issues in certain types of data from their own experiences, creating a resource which might serve useful in new situations. For example, such a system might log found biases in public datasets (particularly when linked to external data) or in whole genres of data, such as Wi-Fi analytics or transport data. Such a system might be a useful starting point for considering issues that may otherwise go undetected, and for supporting low-capacity organisations in their responsible use of analytics. From an HCI perspective though, the design of such a system presents significant challenges. How can often nuanced biases be recorded and communicated both clearly and in such a way that they generalise across applications? How might individuals easily search a system for issues in their own datasets, particularly when they might have a very large number of variables in a combination the system has not seen previously? Making this kind of knowledge accessible to practitioners seems promising, but daunting.

\section{HCI, GDPR and Model Application}

The GDPR, and data protection law in general, was not intended to significantly govern decision-making. Already a strange law in the sense that it poses transparency requirement that applies to the public and private sectors alike, it is also a Frankenstein's monster--style result culminating from the melding of various European law and global principles that preceded it~\cite{fuster2014emergence}. 

\subsection{Modes of Transparency}

While transparency is generally spoken of as a virtue, the causal link between it and better governance is rarely simple or clear. A great deal of focus has been placed on the so-called ``right to an explanation'', where a short paper at a machine learning conference workshop~\cite{goodmanflaxman} gained sudden notoriety, triggering reactions from lawyers and technologists noting that the existence and applicability of such a right was far from simple~\cite{wachter2017right,edwardsveale}. Yet the individualised transparency paradigm has rarely provided much practical use for data subjects in their day-to-day lives (consider the burden of `transparent' privacy policies). Consequently, HCI provides a useful place to start when considering how to make the limited GDPR algorithmic transparency provisions useful governance tools.

There are different places in which algorithmic transparency rights can be found in the GDPR~\cite{wachter2017right}. Each bring different important HCI challenges. 

\paragraph{Meaningful information about the logic of processing} Articles 13--14 oblige data controllers to provide information at the time data is collected around the logics of certain automated decision systems that might be applied to this data. Current regulatory guidance \cite{a29automated} states that there is no obligation to tailor this information to the specific situation of a data subject (other than if they might be part of a vulnerable group, like children, which might need further support to make the information meaningful), although as many provisions in data protection law, the Court may interpret this more broadly or narrowly when challenged. This points to an important HCI challenge in making (or visualising) such general information, but with the potential for specific relevance to individuals. 

\paragraph{Right to be informed} In addition, there is a so-called `right to be informed' of automated decision-making~\cite{wachter2017right}: how might an interface seamlessly flag to users when a potentially legally relevant automated decision is being made? This is made more challenging by the potential for adaptive interfaces or targeted advertising to meet the criteria of a `decision'. In these cases, it is unclear at what point the `decision' is being made. Decisions might be seen in the design process, or adaptive interfaces may be seen as `deciding' which information to provide or withold~\cite{edwardsveale}. Exercise of data protection rights is different in further ways in ambient environments~\cite{edwards2016privacy}, as smart cities and ambient computing may bring significant challenges, if, for example, they are construed as part of decision-making environments. Existing work in HCI has focussed on the difficulties in identifying ``moments of consent'' in ubiquitous computing~\cite{luger2013terms,luger2013aninformed}. Not only is this relevant when consent is the legal basis for an automated decision, but additional consideration will be needed in relation to what equivalent ``moments'' of objection might look like. Given that moments to object likely outnumber moments to consent, this might pose challenges.

\paragraph{A right to an explanation?} An explicit ``right to an explanation'' of specific decisions, after they have happened, sits in a non-binding recital in the GDPR~\cite{wachter2017right}, and thus its applicability and enforceability depends heavily on regulators and the Court. However, there is support for a parallel right in varying forms in certain other laws, such as French administrative law or the Council of Europe Convention 108~\cite{edwardsveale2}, and HCI researchers have already been testing different explanation facilities proposed by machine learning researchers in qualitative and quantitative settings to see how they compare in relation to different notions of procedural justice~\cite{binns2018its}. Further research on explanation facilities in-the-wild would be strongly welcome, given that most explanation facilities to date have focussed on the user of a decision-support system rather than an individual subject to an automated decision.

\subsection{Mitigating Automation Bias}
A key trigger condition for the automated decision-making provisions in the GDPR (art 22)~\cite{edwardsveale} centres on the degree of automation of the process. Significant decisions ``based solely on automated processing'' require at least consent, a contract or a basis in member state law. Recent regulatory guidance indicates that there must be ``meaningful'' human input undertaken by somebody with ``authority and competence'' who does not simply ``routinely apply'' the outputs of the model in order to be able to avoid contestation or challenge~\cite{vealeedwardsa29}. Automation bias has long been of interest to scholars of human factors in computing~\cite{Skitka:1999il,Dzindolet:2003bl} and the GDPR provides two core questions for HCI in this vein. 

Firstly, this setup implies that systems that are expected to outperform humans must always be considered ``solely'' automated~\cite{vealeedwardsa29}. If a decision-making system is expected to legitimately outperform humans it makes meaningful input very difficult. Any routine disagreement would be at best arbitrary and at worst, harmful. This serves as yet another (legal) motivating factor to create systems where human users can augment machine results. Even if this proves difficult, when users contest an automated decision under the GDPR, they have a right to human review. Interfaces need to ensure that even where models may be complex and high-dimensional, decision review systems are rigorous and themselves have ``meaningful'' human input---or else these reviewed decisions are equally open to contestation.

Secondly, how might a data controller or a regulator understand whether systems have ``meaningful'' human input or not, in order to either obey or enforce the law? How might this input be justified and documented in a useful and user-friendly way which could potentially be provided to the subject of the decision? Recent French law does oblige this in some cases: in the case of algorithmically-derived administrative decisions, information should be provided to decision-subjects on the ``the degree and the mode of contribution of the algorithmic processing to the decision-making''~\cite{edwardsveale2}. Purpose-built interfaces and increased knowledge from user studies both seem needed for the aim of promoting meaningful, accountable input.

\subsection{Communicating Envisaged Consequences}

Where significant, automated decision-making using machine learning is expected, the information rights in the GDPR (arts 13--15) provide that a data subject should be provided with the ``envisaged consequences'' of such decision for her. What this means is far from clear. Recent regulatory guidance provides only the example of giving data subject applying for insurance premiums an app to demonstrate the consequences of dangerous driving~\cite{a29automated}. Where users are consenting to complex online personalisation which could potentially bring significant effects to their life, such as content delivery which might lead to echo chambers or ``filter bubbles'', it is unclear how complex ``envisaged consequences'' might be best displayed in order to promote user autonomy and choice. 

\section{Concluding remarks}
 
HCI is well-placed to help enable the regulatory effectiveness of the GDPR in relation to algorithmic fairness and accountability. Here we have touched on different points where governance might come into play---model training and model application---but also different modes of governance. Firstly, HCI might play a role in enabling creative, rigorous, problem solving practices within organisations. Many mechanisms in the GDPR, such as data protection by design and data protection impact assessments, will depend heavily on the communities, practices and technologies that develop around them in different contexts. Secondly, HCI might play a role in enabling controllers do particular tasks better. Here, we discussed the potential for exploratory data analysis tools, such as detecting special category data even when it was not explicitly collected. Finally, it might help data subjects exercise their rights better. It appears especially important to develop new modes and standards for transparency, documentation of human input, and communication of tricky notions such as ``envisaged consequences''.

As the GDPR often defines data controllers' obligations as a function of ``available technologies'' and ``technological developments'', it is explicitly enabled and strengthened by computational systems and practices designed with its varied provisions in mind. Many parts of the HCI community have already been building highly relevant technologies and practices that could be applied in this way. Further developing these with a regulatory focus might be transformative in and of itself---and it is something we believe should be promoted in this field and beyond.
\pagebreak
\bibliographystyle{sigchi}
\bibliography{bib}

\end{document}